\RequirePackage{fix-cm}
\documentclass[twocolumn, a4paper]{article}          
\usepackage{natbib}
\usepackage{amsfonts}
\usepackage{graphicx}
\usepackage{epstopdf}
\usepackage{hyperref}
\usepackage{authblk}
\hypersetup{colorlinks=true,citecolor=black,filecolor=black,linkcolor=black,urlcolor=black}

\title{\bf Linking dynamical and functional properties of
intrinsically bursting neurons}

\author[1]{In\'es Samengo}
\author[1]{Germ\'an Mato}
\affil[1]{Centro At\'omico Ba\-ri\-lo\-che and Instituto Balseiro, San Carlos de Bariloche, Argentina.}

\author[2]{Daniel H. Elijah}
\author[3]{Susanne Schreiber}
\author[2]{Marcelo A. Montemurro \thanks{Email: m.montemurro@manchester.ac.uk}}
\affil[2]{Faculty of Life Sciences, The University of Manchester, United Kingdom.}
\affil[3]{Department of Biology, Humboldt-University Berlin and BCCN Berlin, Germany.}
\date{}
\begin{document}
\renewcommand\Affilfont{\itshape\small}

\maketitle

\hyphenation{con-duc-tance}





\begin{abstract}

Several studies have shown that bursting neurons can encode
information in the number of spikes per burst: As the stimulus
varies, so does the length of individual bursts. The represented
stimuli, however, vary substantially among different sensory
modalities and different neurons. The goal of this paper is to
determine which kind of stimulus features can be encoded in burst
length, and how those features depend on the mathematical properties
of the underlying dynamical system. We show that the initiation and
termination of each burst is triggered by specific stimulus features
whose temporal characteristsics are determined by the types of
bifurcations that initiate and terminate firing in each burst. As
only a few bifurcations are possible, only a restricted number of
encoded features exists. Here we focus specifically on describing
parabolic, square-wave and elliptic bursters. We find that parabolic
bursters, whose firing is initiated and terminated by saddle-node
bifurcations, behave as prototypical integrators: Firing is
triggered by depolarizing stimuli, and lasts for as long as
excitation is prolonged. Elliptic bursters, contrastingly,
constitute prototypical resonators, since both the initiating and
terminating bifurcations possess well-defined oscillation time
scales. Firing is therefore triggered by stimulus stretches of
matching frequency and terminated by a phase-inversion in the
oscillation. The behavior of square-wave bursters is somewhat
intermediate, since they are triggered by a fold bifurcation of
cycles of well-defined frequency but are terminated by a homoclinic
bifurcation lacking an oscillating time scale. These correspondences
show that stimulus selectivity is determined by the type of
bifurcations. By testing several neuron models, we also demonstrate
that additional biological properties that do not modify the
bifurcation structure play a minor role in stimulus encoding.
Moreover, we show that burst-length variability (and thereby, the capacity to transmit information)  depends on a trade-off between the variance of the external signal driving the cell and the strength of the slow internal currents modulating bursts. Thus, our work explicitly links the computational properties
of bursting neurons to the mathematical properties of the underlying
dynamical systems.

\end{abstract}


\section{Introduction}
\label{intro}

In the absence of noise, intrinsically bursting neurons generate
periodic bursts. Periodic activity may serve to control basic
physiological functions, as respiration \citep{Smith1991}, digestion
movements \citep{Selverston1980}, egg laying \citep{Alevizos1991},
or insulin secretion \citep{Meissner1974}. It cannot, however,
transmit information. As input noise increases, bursting responses
become variable \citep{kuske2002,su2004,pedersen2006, chanell2009},
and the number of spikes per burst may fluctuate in time.
Experimental studies performed in different sensory domains have
demonstrated that often, variability in burst length reflects
variability in specific transient features of a time-dependent
stimulus, for example, the orientation of a visual stimulus,
\citep{MartinezConde2002}, the loudness of an auditory stimulus,
\citep{Eyherabide2009}, or the velocity of a somatosensory stimulus
\citep{Arganda2007}. Bursts of different lengths, therefore, are
selective to different stimulus features. This paper aims at
understanding which properties of bursting dynamics determine
stimulus selectivity.

The mapping between burst length and stimulus features is likely to
depend on the biophysical properties of the neuron, such as its
morphology, the types of ionic channels located on the cellular
membrane, their kinetics and their spatial distribution. These
properties may vary continuously in a complex, high-dimensional
space. Yet, from the dynamical point of view, bursting neurons only
display a finite number of qualitatively different behaviors,
depending on the bifurcations governing the underlying dynamical
system \citep{Izhikevich2007}. Previous studies
\citep{Izhikevich2004, Mato2008} have demonstrated that the
computational properties of non-bursting cells, in particular their
selectivity to specific stimulus features,  are determined by the
bifurcation governing the onset of firing. We here hypothesize that
a similar link between dynamical and computational properties can
also be established for bursting neurons.

By combining all the possible bifurcations that can initiate or
terminate spiking, one can in principle construct 120 types of
bursting neurons \citep{Izhikevich2007}. Most of these types,
however, have never been observed experimentally, and some of them
appear only rarely. We therefore focus on the three most ubiquitous
types: parabolic, square-wave and elliptic. In order to explore the
selectivity of each type, we simulate their activity when driven
with a noisy input current and perform a statistical analysis of the
stimulus stretches inducing bursting, obtaining the so-called {\sl
preferred stimuli}. To characterize the input-output mapping of each
type, we search for the amplitude and frequency properties of the
preferred stimuli and interpret those properties in terms of the
underlying bifurcations.

In order to determine whether additional biological details also
shape stimulus selectivity, each type of burster is represented with
three different models, capturing different amounts of biophysical
properties: a minimal normal-form model, a minimal conductance-based
model, and a detailed conductance-based model. The main result is
that stimulus selectivity is mainly shaped by the bifurcation
structure of the burster, and is roughly independent of further
details. The present study thus links the dynamics and the
computational characteristics of bursting neurons.

\section*{Results}

\subsection*{Variability of bursting responses}

When driven with constant currents, intrinsically bursting neurons
tend to generate periodic responses. In the presence of noisy
inputs, however, some variability in the number of spikes per
burst may appear, as well as in the duration of the intra-burst
and the inter-burst intervals. The amount of variability is not
fixed, and example neurons can be found whose responses are
remarkably periodic, or remarkably variable. We begin by
identifying the factors determining the degree of periodicity.

The variables involved in the dynamical description of bursting can
be separated into two classes: fast variables and slow variables
\citep{Rinzel1987}. The former participate in spike generation, and
vary in time scales of the order of 1 ms. The latter vary at a much
slower time scale (tens or hundreds of milliseconds). The evolution
of the membrane potential is governed by equations of the form
\begin{equation}
\frac{{\rm d}V}{{\rm d}t} = f(V) + \sum_i I_{\rm spike}^i + \alpha
\sum_j I_{\rm burst}^j + I_{\rm ext},
\end{equation}
where $f$ is a function of the membrane potential (see Methods),
$I_{\rm spike}^i$ and $I_{\rm burst}^j$ represent the fast and
slow currents respectively, $I_{\rm ext}$ is the external
stimulus, and $\alpha$ is the coupling constant regulating the
degree of coupling between fast and slow variables. The evolution
of the fast and slow currents is determined by additional
variables, governed by additional equations whose functional form
depends on the degree of biological detail (see Methods).

When fast and slow time-scales are well separated, slow variables
define an almost constant current that competes with the external
inputs synapsing on the neuron. Throughout this paper, the external
current is modeled as
\begin{equation}
I_{\rm ext} = I_0 + \sigma \xi(t), \label{e1}
\end{equation}
where $I_0$ is a constant DC, $\xi(t)$ is a stochastic term of zero
mean and unit variance (see Methods) and $\sigma$ represents the
noise intensity.

\begin{figure}
\begin{center}
\includegraphics[width=3.3in]{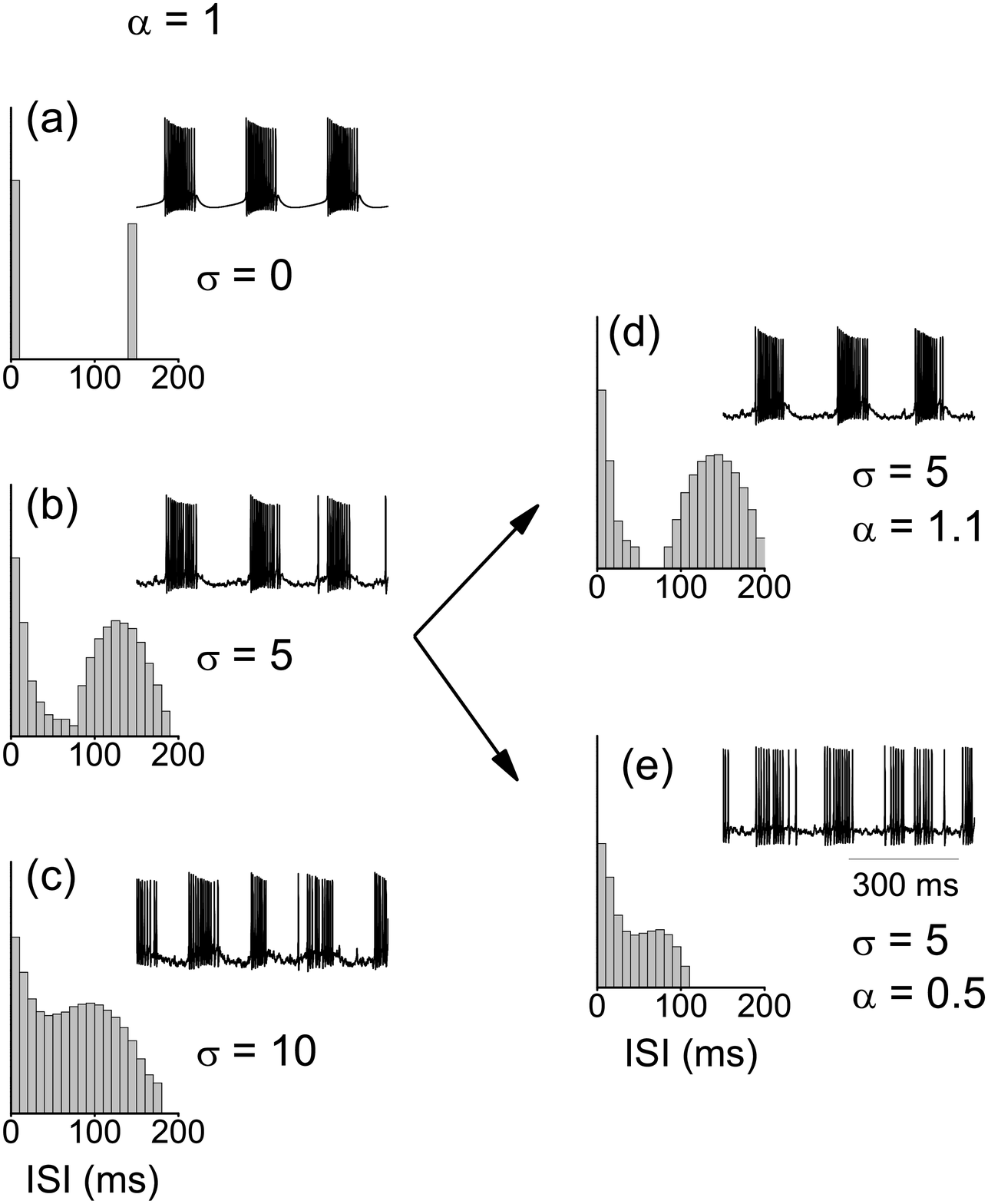}
\end{center}
\caption{ {\bf Periodicity vs. variability of bursting responses}. A
parabolic minimal conductance-based model (see Methods) is
 driven with a noisy input current of standard
deviation $\sigma$ (expressed in $\mu$A ms$^{1/2}$/cm$^2$) and mean
value $I_0$. The magnitude of $I_0$ was chosen to maintain the
average firing rate at 75 Hz. (a) ISI distribution in logarithmic
scale and example voltage trace for the noiseless case. The spike
train is perfectly periodic. (b) and (c): increasing amounts of
noise are employed. The coupling strength is equal as in (a). (d)
and (e): The amount of noise is fixed at 5 $\mu$A ms$^{1/2}$/cm$^2$,
but the coupling strength is increased in (d), and decreased in (e),
resulting in more ordered and disordered responses, respectively.
Noise and coupling strength thus act as opposite forces in
determining response variability. } \label{f1}
\end{figure}

In Figure ~\ref{f1} we show that the amount of variability in
bursting responses is determined by a trade-off between the amount
of noise $\sigma$ in the input signal and the strength $\alpha$ of
the coupling between fast and slow variables. One can therefore
increase the response variability by either increasing the amount of
noise or decreasing the coupling between fast and slow variables.
Panel (a) represents the noiseless situation. The inter-spike
interval (ISI) distribution contains two sharp peaks, the first one
corresponding to the intra-burst intervals and the second one to the
inter-burst silent periods. The voltage trace looks perfectly
regular. As we move to panels (b) and (c), the noise $\sigma$
increases. ISI distributions become wider, and voltage traces appear
more disordered. If the amount of noise remains fixed, however, one
can also vary the degree of disorder by modifying the coupling
strength $\alpha$. The larger the coupling, the larger the
contribution of the slow variables to the evolution of the membrane
voltage. Noise therefore exerts a comparatively weaker influence.
Responses in panels (d) and (e) are calculated with the same
$\sigma$ as in (b). However, increasing $\alpha$ (panel (d))
introduces order, resulting in a sharper ISI distribution and more
regular spiking. Instead, decreasing $\alpha$ (panel (e)) modifies
the responses in a way that is similar to the one obtained with
increased noise (compare with panel (c)). Therefore, coupling
strength and noise act as opposite forces in determining the amount
of variability.

\subsection*{Classification of bursting models in terms of their bifurcations}

In an adiabatic approximation, slow variables can be considered
almost constant in short intervals. At the fast time scales, hence,
slow variables operate as bifurcation parameters of the fast
subsystem. One bifurcation initiates bursting and another one
terminates it. The average values of the fast variables, in turn,
self-consistently modulate the slow subsystem at longer time scales.
The bifurcation structure of different bursting neurons allowed
\citet{Rinzel1987} to classify bursters as parabolic, square-wave or
elliptic.

\begin{figure}
\begin{center}
\includegraphics[width=3.3in]{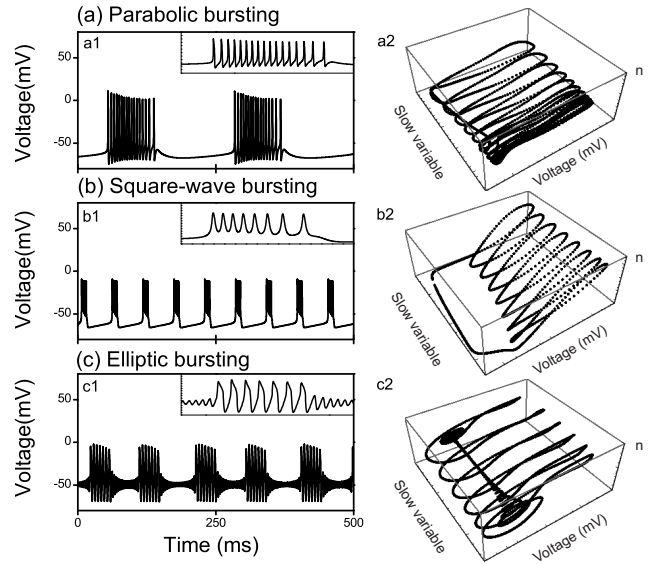}
\end{center}
\caption{ {\bf Properties of different bursting models}. Minimal
conductance-based models were stimulated with a constant external
current, and the voltage trace is displayed (left) as well as a
phase-space trajectory (right). Different types of bursters
exhibit qualitatively different traces. (a): Parabolic burster.
(a1): $I_0 = 5 \mu$A/cm$^2$. (a2): $I_0 = 3 \mu$A/cm$^2$. (b):
Square-wave burster. (b1) and (b2): $I_0 = 6 \mu$A / cm$^2$. C:
Elliptic burster. (c1): $I_0 = 55.5 \mu$A / cm$^2$. (c2): $I_0 =
52 \mu$A / cm$^2$. Insets: amplification of the voltage trace in a
single burst. } \label{f2}
\end{figure}

In parabolic bursters, the initiation and termination of spiking
within each burst is governed by a saddle-node bifurcation on the
invariant circle of the fast subsystem. These transitions are
characterized by large-amplitude and low-frequency oscillations.
As a result, the first and last spike of the burst display ample
voltage deflections. The ISIs, in turn, are modulated throughout
the duration of the burst, displaying a slow-fast-slow sequence.
This modulation gives parabolic bursts their name: The sequence of
ISIs in a burst is a parabolic-shaped function of time. These
features are displayed in Figure~\ref{f2} (a1), for the minimal
conductance-based model. The minimal normal form and the detailed
con-ductance-based models are topologically equivalent.
Figure~\ref{f2} (a2) shows the phase portrait of the parabolic
burster projected on the space defined by the two fast variables,
and one of the two slow variables.

In square-wave bursters, the initiation of spiking is governed by a
fold bifurcation, that is, a saddle-node bifurcation away from the
invariant circle. This transition is characterized by large
amplitude and high-frequency spiking. Burst termination occurs
through a homoclinic bifurcation, distinguished by large amplitude
and low frequency fluctuations. Accordingly, the voltage trace in
Figure~\ref{f2} (b1) shows a deceleration of spiking as the burst
proceeds, without a significant attenuation of the height of voltage
upstrokes. Figure~\ref{f2} (b2) displays a burst in the
3-dimensional phase space.

In elliptic bursters, the initiation of spiking is governed by two
bifurcations, one of them responsible for the loss of stability of
the fixed point representing the resting state, and the other for
the creation of two limit cycles, the stable one representing the
spiking state, and an additional unstable cycle. The fixed point
looses stability through a subcritical Hopf bifurcation, and at
approximately the same critical current value, the two limit cycles
are created through a fold bifurcation of cycles. The combination of
these two bifurcations is called a Bautin bifurcation \citep{Arnold2008}.
Burst termination involve the same two bifurcations occurring
in the reverse direction: The fixed point becomes stable through a
Hopf bifurcation and the limit cycles disappear through a fold
bifurcation of cycles. Elliptic bursts are characterized by
large-amplitude and high-frequency fluctuations. As exemplified in
the inset of Figure~\ref{f2} (c1), the voltage trace exhibits
subthreshold oscillations between bursts, characteristic of Hopf
bifurcations. At spiking onset, the voltage fluctuations suddenly
increase in amplitude and switch to the frequency of the limit cycle
created through the fold bifurcation of cycles. The subthreshold and suprathreshold frequencies are associated with two different attractors, and
therefore, do not necessarily coincide. The termination process is
qualitatively similar. Figure~\ref{f2} (c2) shows the typical
subthreshold oscillations observed in elliptic bursters at the
beginning and at the end of each burst appearing as 3-dimensional
spirals.

\subsection*{Bursting responses obtained with simple stimuli}

Before studying systematically bursting activity in response to
stochastic inputs we briefly survey their behavior when driven with
simple stimuli, as constant or sinusoidal input currents.

\begin{figure*}
\begin{center}
\includegraphics[width=6in]{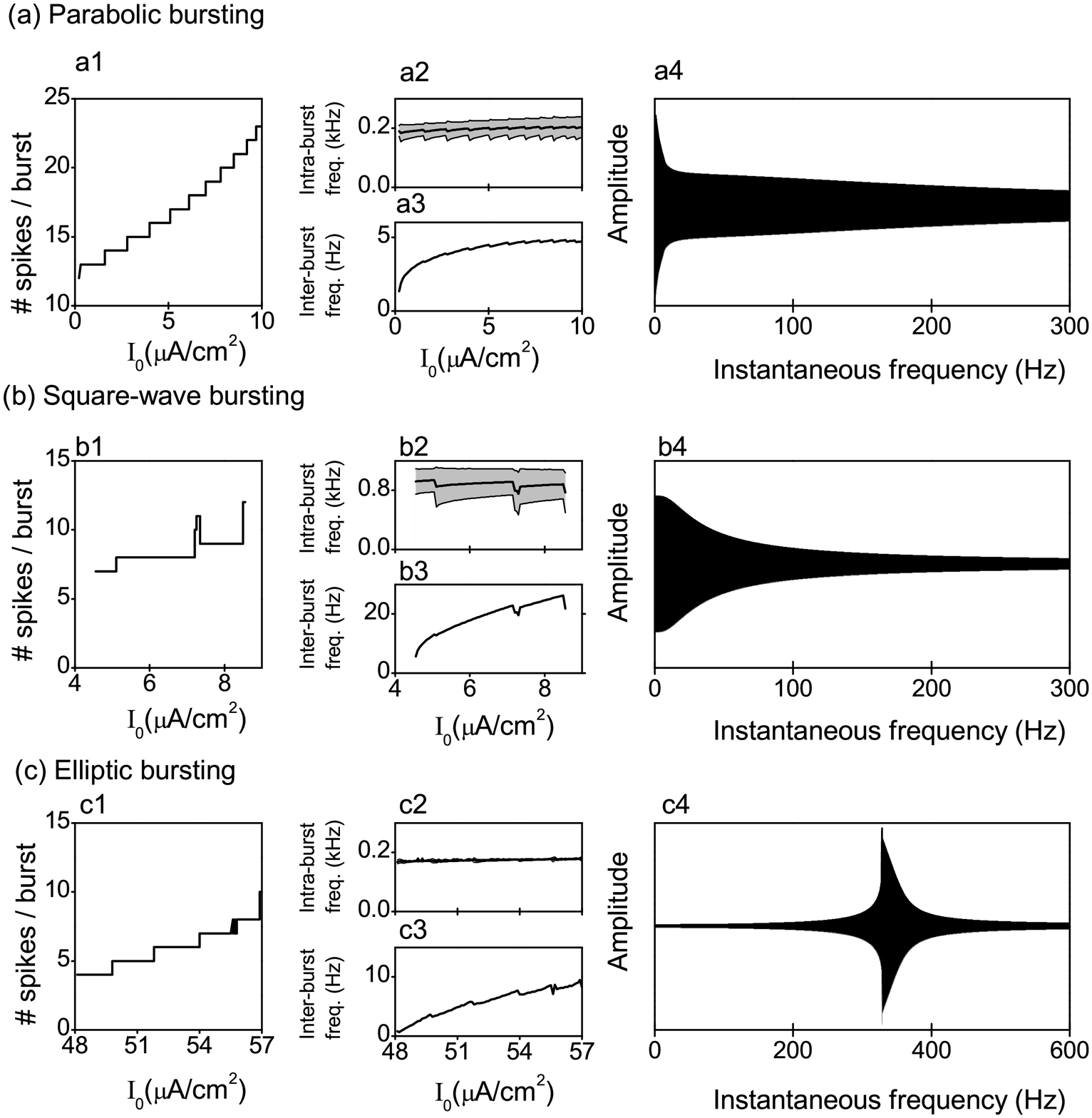}
\end{center}
\caption{ {\bf Response characteristics of different types of
bursting neurons}. Minimal conductance-based models are stimulated
with constant currents, and the number of spikes per burst (a1, b1,
c1) is measured as a function of the input strength. The intra (a2,
b2, c2) and inter-burst (a3, b3, c3) frequencies are calculated as
the inverse of the intra and inter-burst periods. The envelope of
the voltage response to ZAP input currents is an estimation of the
resonance curve for each model. Subthreshold values of $I_0$ have
been used, near to the firing threshold. (a4): $I_0 = 0.1 \mu A /$
cm$^2$, fluctuations of amplitude $I_1 = 0.01 \mu A /$ cm$^2$. (b4):
$I_0 = 4 \mu A /$ cm$^2$, $I_1 = 0.2 \mu A /$ cm$^2$. (c4): $I_0 = 44
\mu A /$ cm$^2$, $I_1 = 0.5 \mu A /$ cm$^2$.} \label{f3}
\end{figure*}

When constant currents are employed ($\sigma$ = 0, in
Equation~\ref{e1}) the number of spikes per burst typically grows
as a function of $I_0$, though not always monotonically
(Figure~\ref{f3} (a1, b1, c1)). In each case, $I_0$ is varied
between a minimum value, below which no spiking occurs, and a
maximum value, beyond which spiking is no longer structured in
bursts, i.e., tonic spiking begins. The mean intra-burst frequency
remains roughly unchanged as $I_0$ varies (panels (a2, b2, c2)). A
constant mean intra-burst frequency, however, does not imply that
all the ISIs inside the burst are equal. Parabolic and square-wave
bursters, in fact, modulate their ISIs throughout the burst, so the
standard deviation of the intra-burst frequency (shaded areas in
Figure~\ref{f3} (a2, b2, c2)) is large, at least, when compared to
elliptic bursters. The dented structure in panels (a2, b2, c2)
corresponds to the steps in panels (a1, b1, c1): When the number of
spikes per burst increases in one unit, an extra spike is packed
into each burst, so the intra-burst ISIs diminish, and so does the
standard deviation. Finally, the inter-burst frequency grows with
$I_0$ (panels (a3, b3, c3)) with only small fluctuations each time
the number of spikes per burst increases.

When driven with subthreshold stimuli, parabolic and square-wave
bursters do not display subthreshold oscillations, since their fixed
points are foci. Elliptic bursters, instead, oscillate at the
frequency associated with the elliptic fixed point (see
Figure~\ref{f2}(c)). Therefore, when driven with subthreshold
sinusoidal stimuli, elliptic bursters may resonate with the input
frequency, as evidenced in the ZAP (impedance (Z) amplitude (A)
profile (P), \citep{Gimbarzevsky1984, Puil1986} depicted in
Figure~\ref{f3}c. Parabolic (a4) and square-wave (b4) bursters have
maximal voltage amplitude for zero-frequency stimulation, whereas
elliptic bursters (c4) have a prominent resonance at 328 Hz.

\subsection*{Classification of bursting models in terms of their degree
of biological detail}

We now turn to the analysis of stimulus selectivity of different
bursting models. Our working hypothesis is that the stimulus
features associated with bursts of a given length depend on the
dynamical properties of the bifurcations associated with the
initiation and termination of bursts. Other biophysical properties
beyond the bifurcation should only weakly affect stimulus
selectivity. To test the hypothesis, we simulated each type of
burster with three different models. The simplest models are
formulated in terms of minimal normal-forms. They involve
dimensionless variables, only loosely related to biophysical
quantities. These models are designed to capture the topological
properties of each bifurcation type in their purest form. As a first
step towards biological realism we also used minimal
conduct-ance-based models, formulated in terms of variables
representing real biological quantities, though not intended to
represent any specific neuron. These models were introduced by
\citet{Izhikevich2007} as the simplest conductance-based models with
the bifurcations of real neurons. Finally, in an attempt to also
include realistic neural models, we simulated detailed
conductance-based models, proposed by \citet{Chay1988}, and further
analyzed by \citet{Bertram1995}. These models represent pancreatic
neurons that, depending on the value of the parameters, can exhibit
parabolic, square-wave, or elliptic bursting.

\subsection*{Stimulus selectivity of different bursting models}

Stimulus selectivity was evaluated by calculating the
burst-triggered average (BTA) of bursts containing a fixed number
of spikes. To do so, we simulated several bursting models,
differentiated by their type (parabolic, square-wave and elliptic)
or by the amount of biophysical detail (minimal normal forms,
minimal conductance-based models, and detailed conductance-based
models). In each case, bursts  were identified by analyzing the ISIs
in the spike train (see Methods). BTAs were calculated by pooling
all the stimulus segments surrounding the generation of bursts of
exactly $n$ spikes and averaging them together. Therefore, each
$n$-BTA represents the average stimulus around bursts of $n$
spikes. We define $t = 0$ as the time of burst initiation.

\begin{figure}
\begin{center}
\includegraphics[width=2.8in]{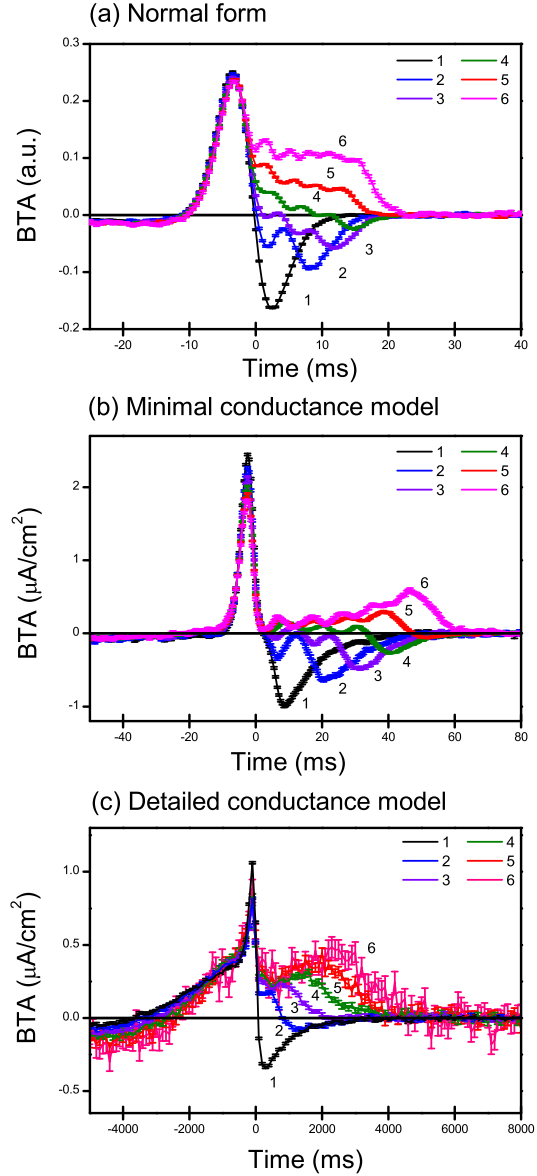}
\end{center}
\caption{ {\bf BTA of parabolic bursters}. Burst onset is at time
zero. (a): Minimal normal form. (b): Minimal conductance-based
model. (c): Detailed conductance-based model. Error bars represent
the error of the mean. Different curves correspond to bursts
containing different number of spikes. In the three models, bursts
are triggered by a strong, depolarizing stimulus upstroke. The
burst is prolonged as long as the excitation is maintained.}
\label{f4}
\end{figure}

In Figure~\ref{f4}, we show the results for the three parabolic
models. Different curves correspond to bursts containing different
number of spikes. The stimulus parameters are given in Tables 1-3.
Although the detailed shape of the BTAs varies with the amount of
biological detail, there are some general trends that are shared by
all three models. In all cases, bursting is preceded by a marked
increase in $I_{\rm ext}$. In the minimal normal form and minimal
conductance models (panels (a) and (b)), bursts are triggered by a
peak that starts rising some 10 ms before burst onset. The detailed
conductance-based model (panel (c)) is governed by much slower time
constants, and the peak begins to become noticeable 1-2 seconds
before burst initiation, an interval 2 orders of magnitude longer
than in the two preceding cases. Hence, the temporal scale of the
structures in the BTA is determined by the details of the model, and
not by the type of bifurcation.

\begin{figure}
\begin{center}
\includegraphics[width=2.8in]{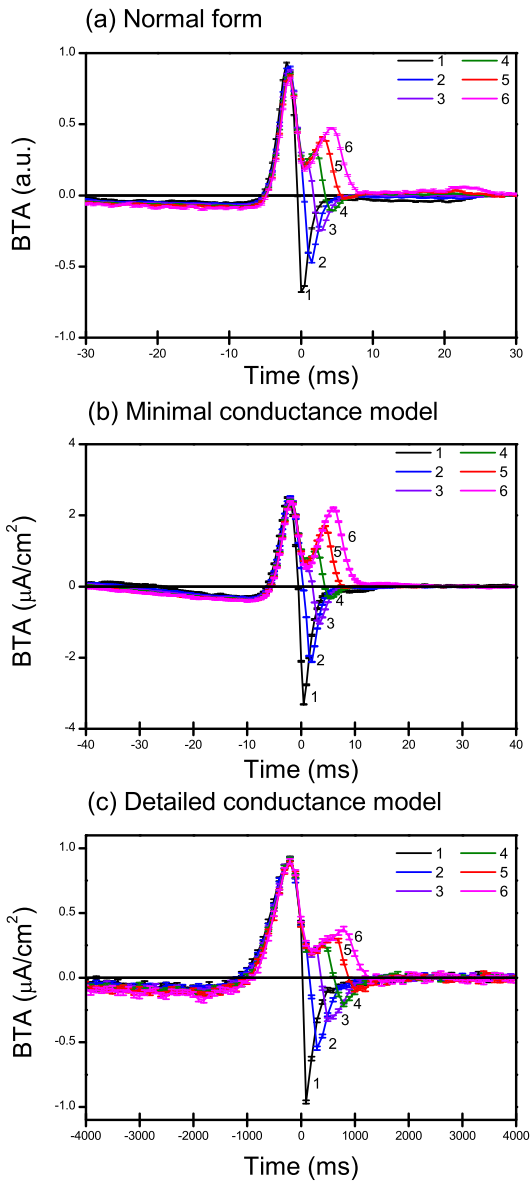}
\end{center}
\caption{ {\bf BTA of square-wave bursters}. Burst onset is at
time zero. Error bars represent the error of the mean. Different
curves correspond to bursts containing different number of spikes.
(a): Minimal normal form. (b): Minimal conductance-based model.
(c): Detailed conductance-based model. In the three models, bursts
are triggered by a small hyperpolarization, followed by a strong,
depolarizing stimulus upstroke. To sustain the burst beyond 3
spikes, a second upstroke is required.} \label{f5}
\end{figure}
\begin{figure}
\begin{center}
\includegraphics[width=2.8in]{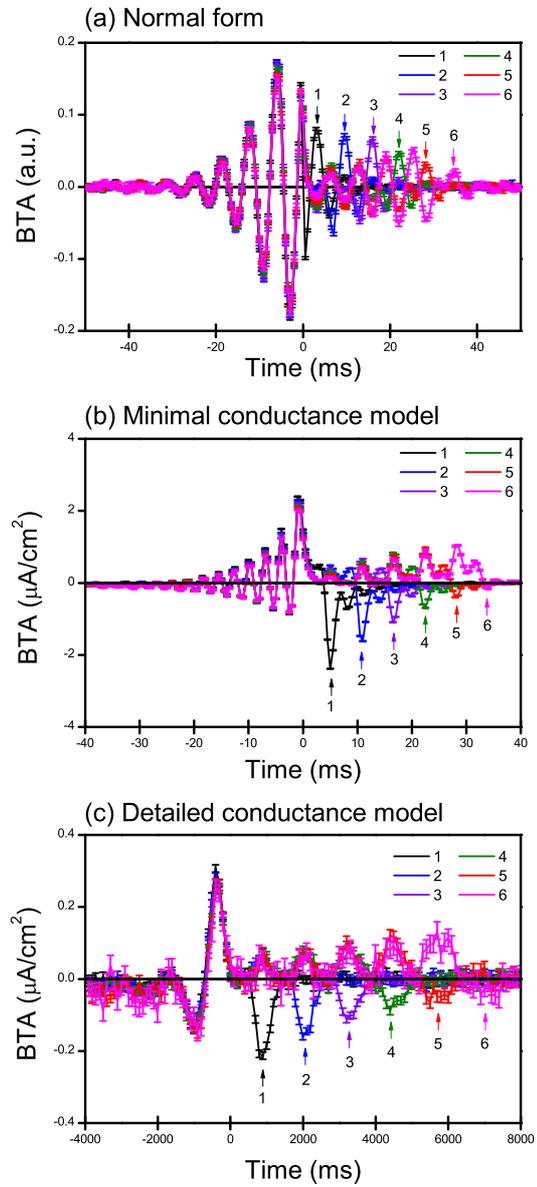}
\end{center}
\caption{ {\bf BTA of elliptic bursters}. Burst onset is at time
zero. Error bars represent the error of the mean. Different curves
correspond to bursts containing different number of spikes. (a):
Minimal normal form. Arrows show the position of the upstroke that
terminates the burst. (b): Minimal conductance-based model. (c):
Detailed conductance-based model. In (b) and (c) arrows show the
position of the downstroke that terminates the burst. In the three
models, bursts are triggered by an oscillating input current of a
well-defined frequency. Bursts are maintained until the input
signal inverts its phase. } \label{f6}
\end{figure}

In the three models, for $t < 0$ all curves overlap. No matter how
many spikes are contained in each burst, bursts are always triggered
by the same type of stimulus deflection. Hence, at the time of burst
initiation, there is no way to predict how many spikes the burst
will contain based on the shape of the stimulus thus far. To
discriminate the stimuli associated with short and long bursts, one
must evaluate the shape of $I_{\rm ext}$ for positive times. Long
bursts appear in response to sustained stimulation beyond burst
initiation. Short bursts are terminated as soon as the stimulus
drops. In the normal form (panel (a)) and in the minimal
conductance-based model (panel (b)), oscillations appear during
sustained stimulation, and including the first large peak, there is
one oscillation per spike in the burst. However, this characteristic
is not a general property of all parabolic models, since it is not
observed in the detailed conductance-based model (panel (c)).

These characteristics may or may not appear in other types of
bursting neurons. In Figure~\ref{f5} we show the BTAs corresponding
to square-wave bursters. Also here bursts are triggered by a sharp,
positive stimulus deflection, and once again, up to the time of
burst onset ($t \le 0$) the shape of the stimulus does not suffice
to predict how long the burst will last. This time, we can also see
a shallow hyperpolarizing phase preceding the burst-triggering peak.
As before, long bursts are associated with prolonged stimulation,
whereas short bursts require a sharp downstroke to terminate. The
shape of both the prolonged stimulation and the downstroke is
remarkably similar across different models (panels (a), (b), and
(c)). However, it differs from the shapes observed in the parabolic
case (Figure~\ref{f4}). Long bursts in square-wave bursters require
stimuli containing two separate peaks. In order to sustain spiking
beyond the mean burst length (3 spikes per burst, for the mean
currents used in Figure~\ref{f5}), an additional positive derivative
in the current is required. This stimulus upstroke marks the
initiation of the second peak in the BTA. In the minimal normal
form, the second peak begins 0.5 ms after burst onset, and in the
other two models, at 1 and 200 ms. In the three models, the timing
between burst initiation and the beginning of the second peak
coincides with the period of the spiking limit cycle at burst onset.
Hence, to get long bursts we need a 2-peak oscillation in the
stimulus whose frequency coincides with the frequency of the limit
cycle of the fold bifurcation initiating spiking. The second peak in
the stimulus therefore begins to be noticeable at the time when the
second spike of the burst is fired.

Elliptic bursting is markedly different from the previous two. In
Figure~\ref{f6} we see that the BTAs of the three elliptic models
contain a pronounced oscillatory structure. The frequency of the
oscillations remains roughly constant throughout the whole of the
BTA in the normal form model ((a)). In the conductance-based
models (panels (b) and (c)), the frequency diminishes after burst
initiation. In all three cases, the frequency before burst
initiation coincides with the frequency associated with the spiral
fixed point, defined by the imaginary part of the eigenvalue loosing
stability. The frequency after burst initiation coincides with the
frequency of the spiking limit cycle. The two frequencies coincide
in the case of the normal form (a), but do not coincide in the
conductance-based models (b) and (c).

Burst initiation is triggered by gradually amplifying stimulus
oscillations. Once again, the BTAs corresponding to bursts of
different duration coincide for $t \le 0$. Burst termination may be
due to a marked upstroke (panel (a)) or downstroke (panels (b) and
(c)), as indicated by the arrows in Figure~\ref{f6}. In all cases,
the terminating feature (be it an upstroke or a downstroke) is out
of phase with respect to the ongoing oscillation. Hence, burst
termination is governed by a sudden phase inversion in the stimulus.
The fact that the phase inversion may appear either as a peak or a
trough implies that the extinction of the burst does not depend on
the sign of the stimulus deflection, but rather, on the phase
inversion.

\subsection*{Dependence on simulation parameters}

\begin{figure*}
\begin{center}
\includegraphics[width=5in]{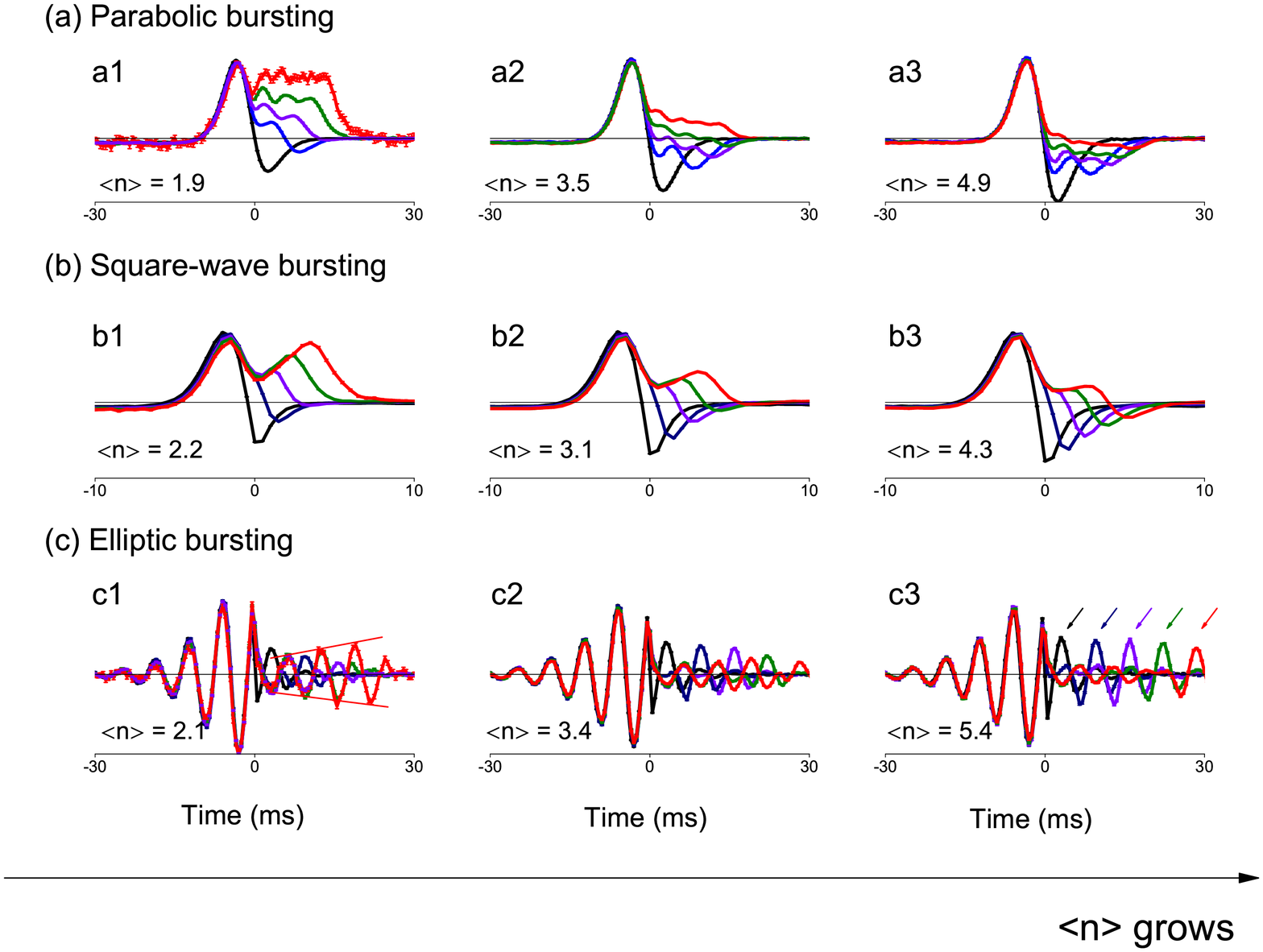}
\end{center}
\caption{ {\bf BTAs for different coupling strengths $\alpha$ in
normal-form models}. Burst onset is at time zero. Error bars
represent the error of the mean, sometimes too small to be visible.
Different curves correspond to bursts containing different number of
spikes ranging from $n = 1$ to $n = 5$ with the same color code as
in Figures~\ref{f4}-\ref{f6}. (a) Parabolic bursting. a1: $\alpha =
0.91$. a2: $\alpha = 1$. a3: $\alpha = 1.05$. (b) Square-wave
bursting. b1: $\alpha = 1.81$. b2: $\alpha = 1$. b3: $\alpha =
0.55$. (c) Elliptic bursting. c1: $\alpha = 1.6$. c2: $\alpha = 1$.
c3: $\alpha = 0.6$.} \label{f7}
\end{figure*}

Stimulus selectivity varies in a systematic fashion when the
baseline current $I_0$, the amount of noise $\sigma$, and the
coupling strength $\alpha$ are modified. In Figure~\ref{f7} we
display the BTAs of the normal-form models previously shown in
Figures~\ref{f4}-\ref{f6} but now for different coupling strengths
$\alpha$.

In parabolic and square-wave bursters, as $\alpha$ changes, the
family of BTAs obtained for different $n$ values shifts either
upward or downward. The direction of the shift is determined by
the mean number of spikes per burst $\langle n \rangle$, which in
turn, is a function of $\alpha$. For a given $\alpha$, the mean
duration of each burst is $\langle n \rangle$. Bursts with $n >
\langle n \rangle$ (panels (a1) and (b1)) require sustained
excitation in order to last longer than the typical length
$\langle n \rangle$. Bursts with $n < \langle n \rangle$ (panels
(a3) and (b3)), instead, require a prompt downstroke to terminate
firing rapidly, before the length $\langle n \rangle$ is reached.
In panels a2 and b2, $\langle n \rangle \approx 3$, so some of the
curves display downstrokes (those with $n < \langle n \rangle$)
and others sustained excitation (those with $n > \langle n
\rangle$).

The relationship between $\alpha$ and $\langle n \rangle$ is not
universal, i.e., different models show different trends. For
example, increasing $\alpha$ results in a larger $\langle n \rangle$
in the square-wave model of Figure~\ref{f7} (b), and a smaller
$\langle n \rangle$ in the parabolic case of panel (a). Yet, in
conductance-based models different trends may be found. The key
factor determining whether $\langle n \rangle$ grows or diminishes
with $\alpha$ is the sign of the average value of the slow current.
The current $I_{\rm burst}$ exerts an influence on the fast
sub-system that may be excitatory, inhibitory or neutral on average.
In the parabolic model of Figure~\ref{f7} (a) $I_{\rm burst}$ is
excitatory. In the square-wave model (panel b), $I_{\rm burst}$ is
inhibitory. Yet other models sharing the same bifurcations can be
engineered with mean slow currents of opposite sign. Therefore, the
relationship between $\alpha$ and $\langle n \rangle$ is not a
universal property of all parabolic or square-wave models. Once the
relationship is given, however, stimulus selectivity varies
systematically as a function of $\alpha$.

Elliptic bursters behave in the same way, though the effect is less
visible by naked eye, because prolongation and termination features
are not easily separated visually. After burst onset, in-phase
oscillations constitute prolongation features. The last oscillation,
appearing with inverted phase is a terminating feature. When $n >
\langle n \rangle$, bursts need to be prolonged beyond their natural
duration. The prolongation becomes increasingly difficult as time
goes by, so the oscillations sustaining bursts during positive times
increase in amplitude as the burst proceeds (marked with two lines
in Figure~\ref{f7} c1 for $n$ = 5). The terminating feature is
instead small. When $n < \langle n \rangle$, bursts need to be
terminated before they reach their natural duration. The termination
signal must be particularly strong when the natural duration is
large (large $\langle n \rangle$). Therefore, the out-of-phase
terminating feature in c3 (marked with  arrows, one for each
$n$) has a larger amplitude than in c1.

Stimulus selectivity also varies when other parameters are modified,
as for example the stimulus mean $I_0$ or the standard deviation
$\sigma$. In these cases, again, the critical factor is how these
variations affect $\langle n \rangle$. Changes in $\langle n
\rangle$ originated in changes in $I_0$ produce BTAs that, when
arranged in increasing value of $\langle n \rangle$, are in all
similar to the ones depicted in Figure~\ref{f7}. Changes in $\sigma$
also give rise to the same phenomena, with one additional effect:
For large $\sigma$, BTAs have less memory of events occurring in the
distant past or distant future. As a consequence, BTAs look the same
as in Figure~\ref{f7}, but with a damped envelope that gradually
decays for positive and negative times.

\subsection*{Independent components in stimulus selectivity}

The BTA is the average stimulus preceding a burst. When a single
burst is generated, the triggering stimulus typically differs from
the BTA up to a certain degree. Therefore, one can calculate not
only the mean stimulus of each $n$, but also the variability around
the mean. When studying a collection of scalar quantities,
variability is captured by the standard deviation. With vectorial
quantities, a single number does not suffice to describe
variability, since there are many possible directions in which
fluctuations may appear. Covariance analysis provides a tool for
detecting the stimulus directions where uncorrelated variations are
observed. Noticeably, variations are particularly large, or
particularly small in only a few directions. These are the so-called
{\em relevant directions}: Stimulus directions where the standard
deviation of the burst-triggering ensemble is either larger or
smaller than in the prior stimulus (see Methods). Covariance
analysis also provides a systematic procedure to reveal these
relevant directions \citep{samengo2013}. Here, the analysis is
carried out for each intra-burst spike count $n$. The stimulus
segments preceding bursts containing exactly $n$ spikes are
identified, and their temporal correlations are captured by the
covariance matrix. The eigenvectors of this matrix represent the
relevant features, and the corresponding eigenvalues provide a
quantitative measure of the variance of the data in these directions
(see Methods). Typically, most burst-eliciting stimulus directions
have approximately the same variance as the total stimulus, and only
a few depart noticeably from the general trend. The eigenvectors
associated with outlier eigenvalues constitute the relevant stimulus
directions.

\begin{figure}
\begin{center}
\includegraphics[width=3.3in]{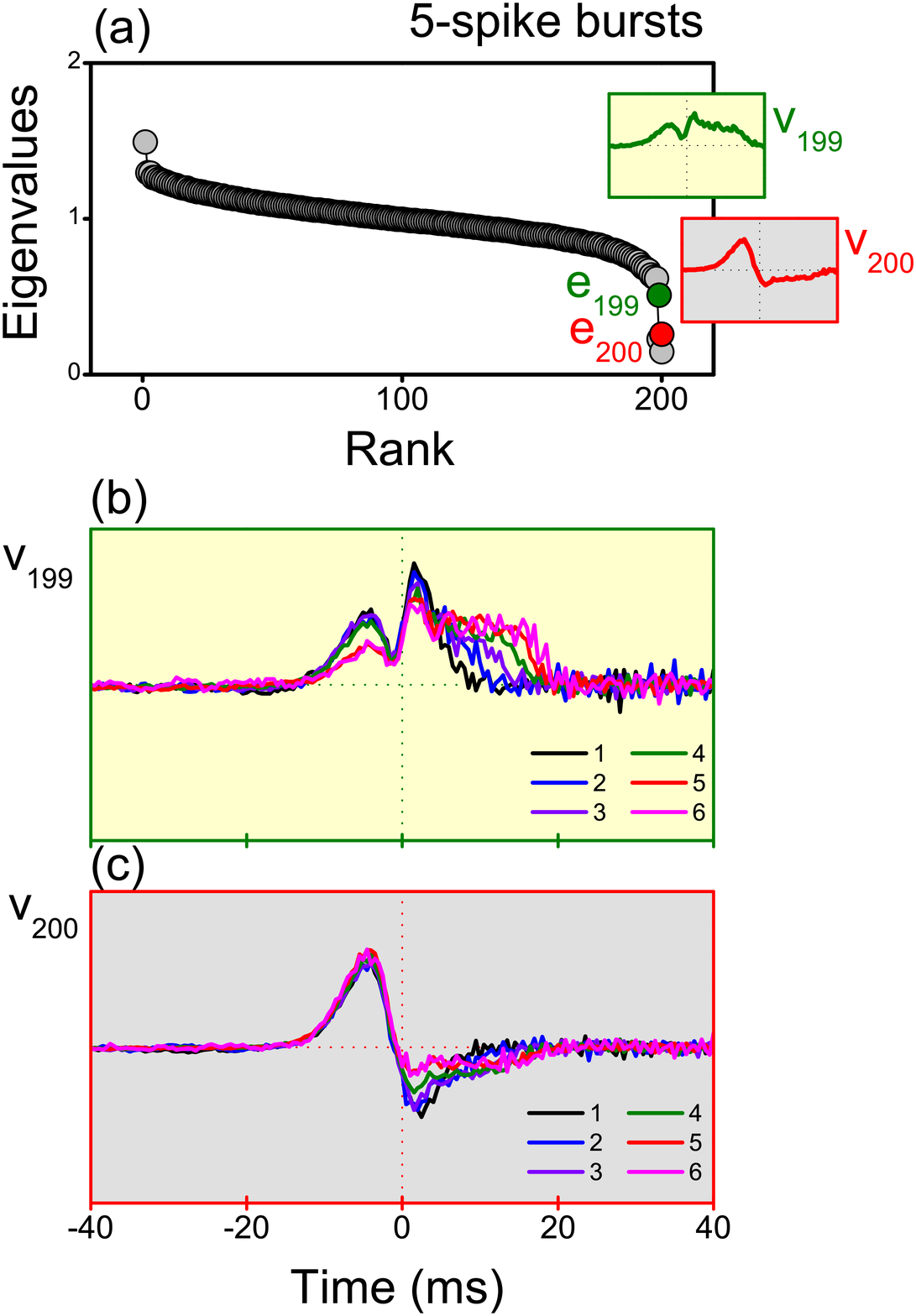}
\end{center}
\caption{ {\bf Covariance analysis of a parabolic burster}. Data
obtained by simulation of the minimal normal-form model. (a):
Example spectrum of eigenvalues for stimulus segments triggering
bursts containing 5 spikes. The two most clear outliers ($e_{199}$
and $e_{200}$) are marked. Insets: eigenvector corresponding to each
eigenvalue. (b-c): Eigenvectors associated with outlier eigenvalues,
for bursts containing $n$ spikes (one curve for each value of $n$).
Burst onset is at time zero. Eigenvector ${\rm v}_{200}$ only
contains stimulus features initiating the burst, whereas ${\rm
v}_{199}$ mainly describes structures needed to sustain and
terminate the burst. } \label{f8}
\end{figure}

In Figure~\ref{f8}, we show the eigenvalues and eigenvectors of
the normal form of a parabolic burster. The two most prominent
outlier eigenvalues are ${\rm e}_{199}$ and ${\rm e}_{200}$, and
both of them have decreased variance. In panels (b-c) we see the
eigenvectors associated with each of the outlier eigenvalues, for
bursts containing different numbers of spikes.

Eigenvector ${\bf v}_{200}$ is the most significant relevant
feature, since its eigenvalue has the largest deviation from
unity. Its shape strongly resembles the BTA (Figure~\ref{f4}),
except for the lack of a plateau at positive times, for long
bursts. There is virtually no difference in the shape of the
vectors corresponding to bursts containing different numbers of
spikes. Variations in this direction capture the fluctuations in
the stimulus deflection triggering bursts. Eigenvector ${\bf v}_1$
is therefore involved in shaping burst initiation, but it contains
no features related to burst termination.

Parabolic bursting lasts for as long as the stimulus depolarizes
the cell (Figure~\ref{f4}). That is, longer bursts are associated
with more prolonged stimulation. Eigenvector ${\bf v}_{199}$
captures these differences: In Figure~\ref{f8}(b), we see that
bursts containing a single spike are briefly stimulated, whereas
6-spikes bursts are driven for almost 20 ms after burst onset.
Therefore, eigenvector ${\bf v}_{199}$ crucially captures burst
termination.

Since burst initiation and burst termination are described by two
different eigenvectors (${\bf v}_{200}$ and ${\bf v}_{199}$,
respectively), these two processes are uncorrelated: Fluctuations in
the features associated with burst initiation are not correlated
with fluctuations responsible for burst termination.

\begin{figure}
\begin{center}
\includegraphics[width=3.3in]{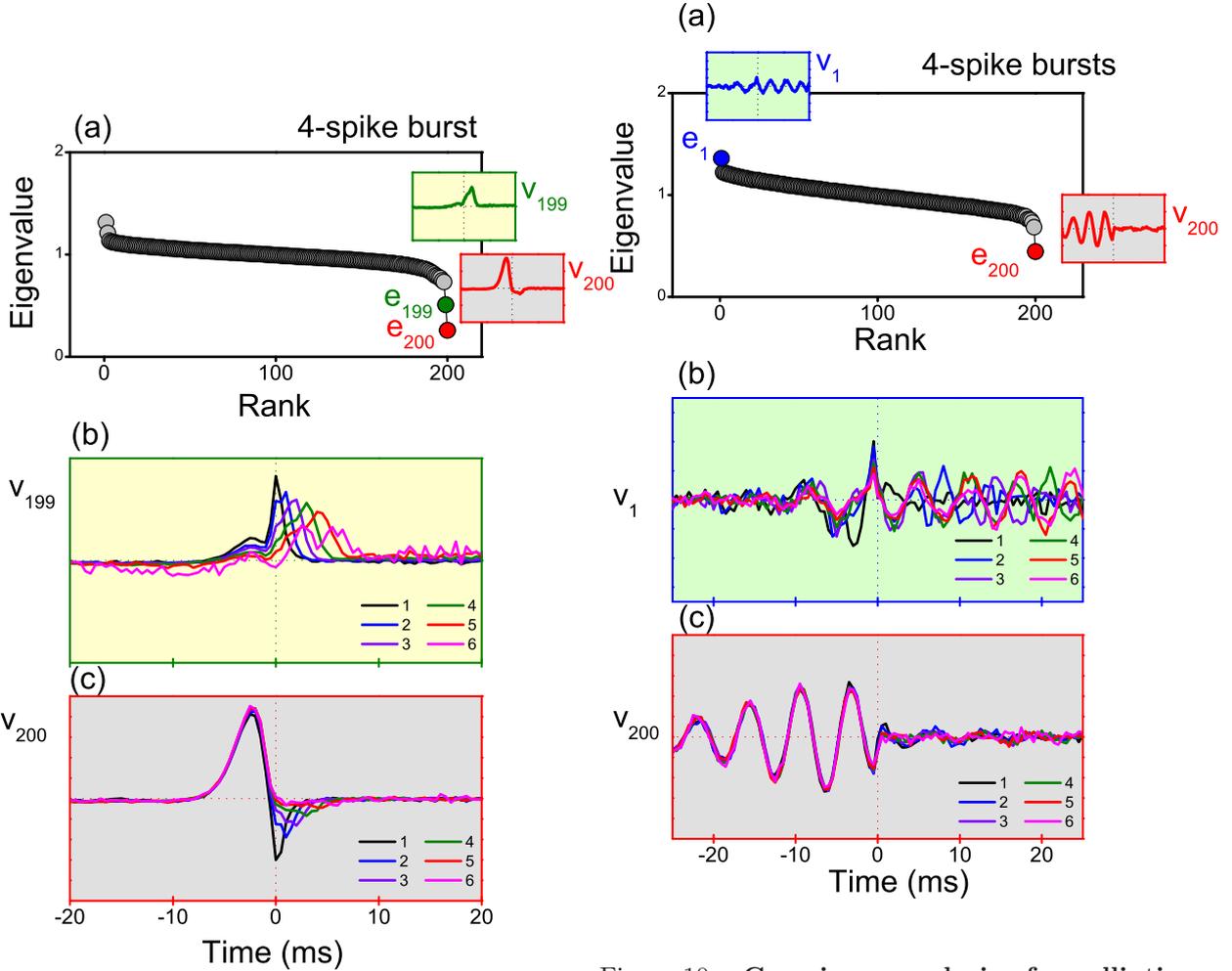}
\end{center}
\caption{ {\bf Covariance analysis of a square-wave burster}. Data
obtained by simulation of the minimal normal-form model. (a):
Example spectrum of eigenvalues for stimulus segments triggering
bursts containing 4 spikes. Two outliers ($e_{199}$ and $e_{200}$)
are visible. Insets: eigenvector corresponding to each eigenvalue.
(b-c): Eigenvectors associated with outlier eigenvalues for bursts
containing $n$ spikes (one curve for each value of $n$). Burst onset
is at time zero. Eigenvector ${\rm v}_{200}$ contains stimulus
features initiating the burst, whereas ${\rm v}_{199}$ describes the
structures needed to sustain and terminate the burst.} \label{f9}
\end{figure}

\begin{figure}[h!]
\begin{center}
\includegraphics[width=3.3in]{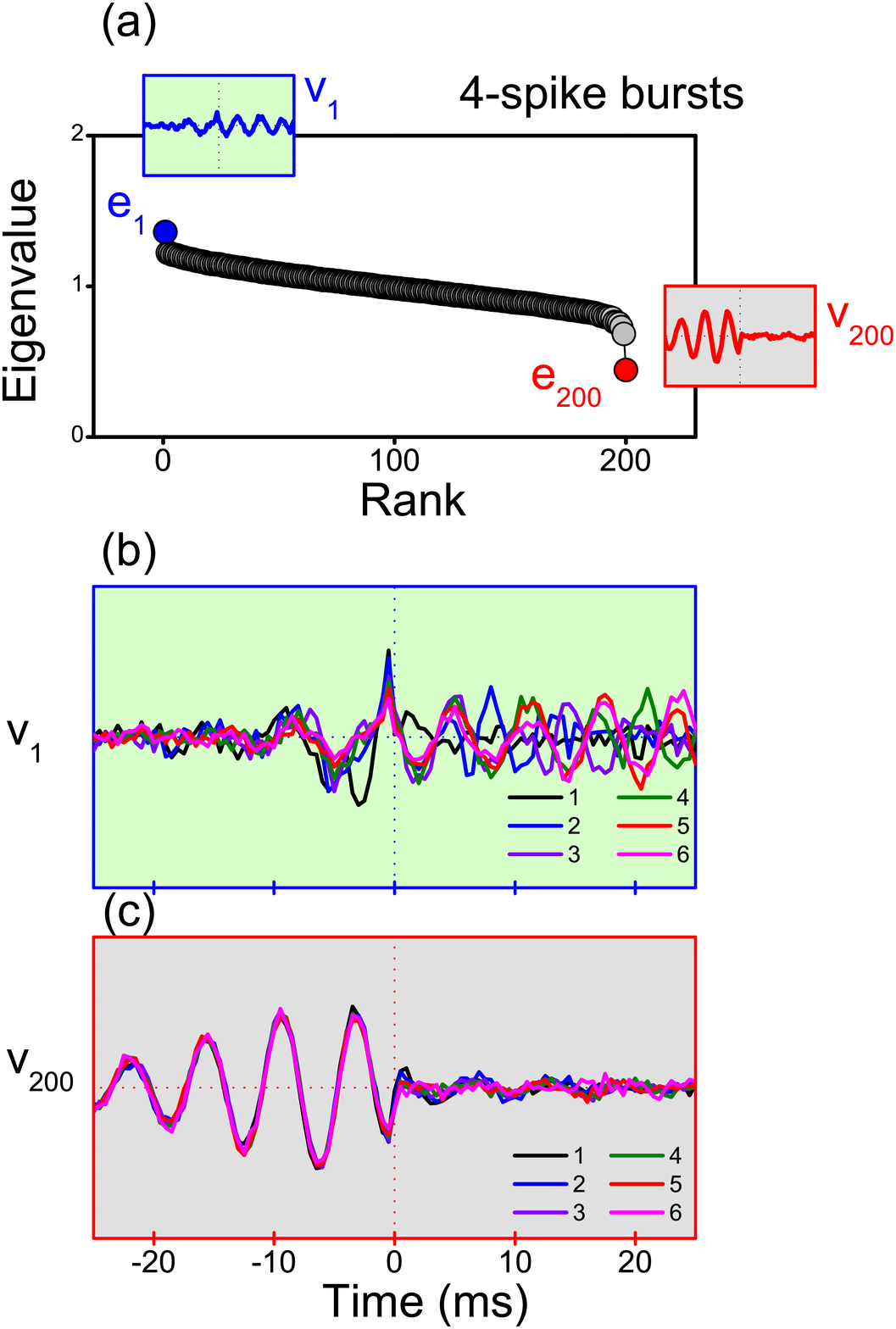}
\end{center}
\caption{ {\bf Covariance analysis of an elliptic burster}. Data
obtained by simulation of the minimal normal-form model. (a):
Example spectrum of eigenvalues, for stimulus segments triggering
bursts containing 4 spikes. Two outliers (${\rm v}_1$ and ${\rm
v}_{200}$) are visible. Insets: eigenvectors corresponding to each
eigenvalue. (b, c): Eigenvectors associated with outlier
eigenvalues, for bursts containing $n$ spikes (one curve for each
value of $n$). Burst onset is at time zero. Eigenvector ${\rm
v}_{200}$ contains stimulus features initiating the burst, whereas
${\rm v}_{1}$ describes the structures needed to sustain and
terminate the burst.} \label{f10}
\end{figure}

A similar situation is found for square-wave and elliptic
bursters, as shown in Figures~\ref{f9} and \ref{f10}. In both
cases, eigenvector ${\bf v}_{200}$ describes the shape of the
stimulus features initiating the burst. In the case of a
square-wave burster (Figure~\ref{f9}(c)), this feature is a
depolarizing upstroke, whereas in the elliptic case
(Figure~\ref{f10}(c)), bursting is initiated by an oscillation
whose frequency coincides with the subthreshold resonance
frequency of the cell. The remaining eigenvector (${\bf v}_{199}$
in the square-wave burster, and ${\bf v}_{1}$ in the elliptic
case) represents the stimulus features terminating the burst.

When working with conductance-based models, similar results are
obtained: Stimulus features responsible for burst initiation and
termination tend to appear in separate eigenvectors. The
eigenvectors responsible for burst initiation remain unchanged when
$n$ is varied. Those responsible for burst prolongation or
termination, instead, mark the differences in stimulus selectivity
of different $n$ values.

\section*{Discussion}

Initial studies on bursting neurons showed that they fire in a
stereotyped fashion, and that regular bursting appears spontaneously
even in culture tissue \citep{ChagnacAmitai1989, Silva1991,
Steriade1991, Agmon1992}. Bursting was thus assumed to serve as a
pacemaker signal in drowsy or anesthetized states
\citep{Steriade1991, Steriade1993}, but could not encode transient
information, since a perfectly periodic signal has zero information
rate. More recent studies, however, have shown that the number of
spikes per burst may encode specific properties of the input
stimulus. In sensory systems, these include the visual system of
mammals \citep{DeBusk1997, MartinezConde2002}, the auditory system
of insects \citep{Eyherabide2008, Eyherabide2009, Marsat2010}, the
somatosensory system of leech \citep{Arganda2007} and the olfactory
system of rodents \citep{Cang2003}. Theoretical studies have shown
that the number of spikes per bursts encode the slope
\citep{Kepecs2002} or the phase \citep{Samengo2010} of the input
current driving a cell.

Our analysis shows that these two behaviors constitute two extremes of
a continuum phenomenon. Depending on the trade-off between the
amount of noise in the external signal and the internal
burst-related currents, neurons vary their coding capacity in a
graded manner. When the external signal is constant, spike trains
are perfectly regular. Neurons generate periodic bursts whose
duration is determined by the oscillatory modulation produced by the
slow variables. Since all bursts contain the same number of spikes,
no information is encoded in burst duration. As fluctuations are
incorporated to the external stimulus, bursts become irregular (see
Figure~\ref{f1}). The degree of irregularity is determined by a
trade-off beetween the amount of noise in the input and the coupling
between fast and slow variables.

However, irregularity may or may not be informative. Bursts
containing different number of spikes are only informative if they
are selectively triggered by stimuli with specific properties. In
order to assess whether there is a correspondence between burst
length and stimulus attributes we calculated BTAs associated with
different $n$ values. We showed that the stimulus features
associated with burst firing depend on the type of bifurcations
initiating and terminating spiking in each burst, and are roughly
independent of additional biological properties
(Figures~\ref{f4}-\ref{f6}).

The three explored models can be arranged along a
re\-so\-na\-tor-integrator dimension. Elliptic bursters behave as
emblematic resonators, as seen from their sharp subthreshold
resonance and their precisely tuned intra-burst frequency
(Figure~\ref{f2}). Correspondingly, stimuli triggering, prolonging
and terminating elliptic bursts have strong oscillatory components.
Parabolic bursters behave diametrically opposite: They are
archetypical integrators. Their two bifurcations have slow onsets
and lack a characteristic time scale. Accordingly, stimuli
triggering parabolic bursts are depolarizing currents that need to
be sustained for as long as the burst lasts. Square-wave bursters
lie somewhere in between. The initiating bifurcation has a well
defined frequency, but not the terminating one. Accordingly, stimuli
triggering quadratic bursting have mixed characteristics: BTAs
associated with long bursts contain two peaks, and the time interval
between burst onset and the initiation of the second peak coincides
with the frequency of the firing limit cycle at burst onset.

Covariance analysis showed that stimulus features triggering burst
initiation appear in separate eigenvectors as those terminating
bursts, implying that the two processes are uncorrelated. Moreover,
stimulus features responsible for burst initiation do not depend on
burst duration. At the time of burst onset, therefore, one cannot
predict the duration of the burst based on the evolution of the
stimulus thus far. Therefore, the generation of a burst --regardless of its length--
 reduces the uncertainty of the stimuli preceding burst
generation, and thereby carries information about past stimuli. The
length of the burst, in turn, reduces the uncertainty of the
evolution of the stimulus during the duration of the burst, and
thereby carries information about the events that take place between
burst initiation and burst termination.

Parameters as the DC stimulus component $I_0$, the standard
deviation $\sigma$ and the coupling constant $\alpha$ affect the
correspondence between $n$ values and stimulus attributes. The
changes, however, never mix different bursting models, that is,
elliptic bursters are selective to oscillatory stimuli no matter the
value of the parameters, and similar conclusions can be drawn for
parabolic and square-wave bursters. When parameters are modified,
the mean number of spikes per burst $\langle n \rangle$ varies. If
$\langle n \rangle$ increases, it becomes more difficult to obtain
short bursts, and therefore, the few short bursts that appear occur
in response to markedly pronounced terminating features. If several
parameters are changed simultaneously in such a way as to leave
$\langle n \rangle$ constant, stimulus selectivity remains roughly
unchanged.

Our results imply that in a given experiment, one may apply reverse
correlation techniques to obtain BTAs and eigenvectors for different
burst lengths, and from their shape, deduce the bifurcations
governing the underlying dynamical system. Thereby, intrinsic
properties of the recorded cells can be identified, even with
extracellular measurements, where no information about the temporal
evolution of the subthreshold activity is available. Care should be
taken, however, to fulfill the following conditions.
\begin{itemize}
\item[-] The experiment should function in a range of mean resting
potentials where the recorded cell bursts intrinsically. In some
experiments, all observed bursts are driven by transient stimulus
fluctuations, and the neuron stops bursting as soon as a constant
stimulus is applied. Those cases are not described by the present
theory, because the underlying dynamical system contains a single
bifurcation: the one describing the firing threshold. Such
experiments should be analyzed with the theory developed for tonic
neurons \citep{Mato2008}.

\item[-] The random stimulus should be appropriate for reverse correlation.
In particular, it should have a high enough cutoff frequency, so
that the missing high frequency components be irrelevant to neuronal
dynamics, that is, when present, they be filtered out by the
capacitive properties of the cell membrane. When this condition is
not fulfilled, BTAs contain oscillations that do not represent
intrinsic neuronal properties, but prominent stimulus frequencies,
typically, the cutoff frequency. In order to avoid such artifacts,
the required corrections for applying reverse-correlation techniques
to colored stimuli should be made \citep{samengo2013}. Even so, it
should be noticed that those frequencies that are absent in the
stimulus, are also absent in the obtained BTAs and eigenvectors,
perhaps hindering the identification of prominent frequencies in the
BTAs of elliptic bursters.

\item[-] BTAs and eigenvectors should extend to positive times long
enough as to ensure that all terminating features are contained.
They should also be discriminated by the number of spikes per burst
so that terminating features can be seen gradually displaced in
time, for longer bursts.

\item[-] Noise should not be too large as to occlude the temporal
evolution of BTAs and eigenvectors for positive and negative times.
Recall that as noise increases, both BTAs and eigenvectors decay
rapidly in time, so the features extending to the past or the future
may not be discernible. Oscillations, for example, may damp out
before a full cycle is completed. Since it may not be easy to
decide, in a given experiment, which noise levels are appropriate,
in doubtful situations, different runs using several noise levels
are recommended.

\end{itemize}
If these conditions are met, and assuming that the recorded cell
produces either parabolic, square-wave or elliptic bursts, one may
speculate which of the three alternatives best describes the data.
Elliptic bursters are the best candidates for strongly oscillating
BTAs with out-of-phase terminating features. Parabolic bursters
constitute the best choice for purely depolarizing BTAs that last
for as long as firing is sustained. Square-wave bursters are the
appropriate choice when the upstroke triggering bursts is preceded
by a shallow hyperpolarization and long bursts are sustained by a
two-peak stimulus structure.

Altogether, the function of bursters of the three major bifurcation
types range from pure pacemaking to the encoding of information in
the spike count per burst. In particular, burst onset and
termination are largely uncorrelated processes and can hence
independently modify the trade-off between pacemaking and
information transfer. Moreover, the initiating and terminating
features are determined by the specific bifurcations underlying
burst initiation and termination. This implies that key
computational aspects of bursting do not depend on the particular
biological details of the neuron but rather on the basic topological
structure of the underlying dynamical system. Physiologically,
bursters and the network they are embedded in have a wide repertoire
of possibilities to regulate their function. In particular, our
study suggests they can do so via changes in the coupling constant
$\alpha$, such as adaptation processes or short-term plasticity, as
well as via changes to the effective input to the bursting neuron,
such as the degree of synchronization in presynaptic neurons.

\section*{Methods}
\subsection*{Burst identification}

\begin{figure}
\begin{center}
\includegraphics[width=3.3in]{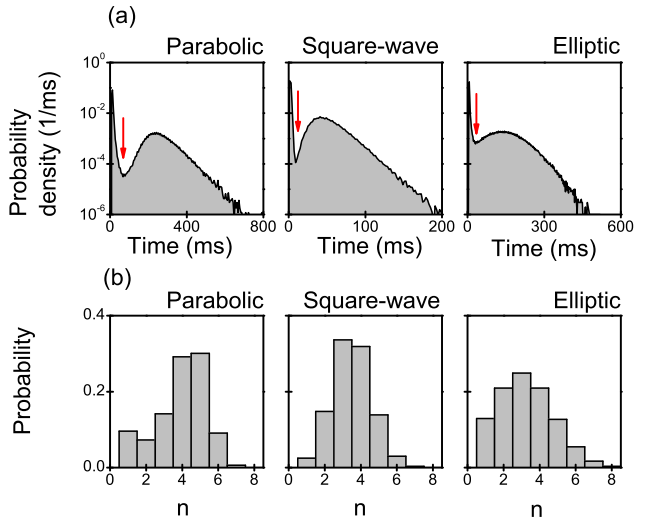}
\end{center}
\caption{ {\bf Statistical properties of bursting neurons}. Example
simulation of the minimal conductance-based models. Similar results
are obtained with the normal forms, or the detailed
conductance-based models. (a): ISI probability densities, with the
local minimum defining the threshold ISI marked by arrows.
Smaller ISIs correspond to intra-burst intervals, and longer ISIs,
to inter-burst intervals. (b): Probability of finding a burst
containing $n$ spikes, for different types of bursting neurons.}
\label{f12}
\end{figure}

Spike trains are parsed into sequences of bursts. Two consecutive
spikes are assigned to the same burst or to different bursts,
depending on their ISI. If the ISI is larger than a pre-defined
threshold, the two spikes are assigned to different bursts.
Otherwise, they are taken as part of the same burst. The threshold
ISI is defined as the one where the ISI distribution reaches a local
minimum (arrows, in the example of Figure~\ref{f12} (a)). In all the
simulations, the ISI distribution has a bimodal structure. This
structure allows us to define the limiting ISI without ambiguity.
Once spike trains are parsed into bursts, a distribution of $n$
values is obtained (Figure~\ref{f12} (b)). The mean and variance of
this distribution depend on $\alpha$, $I_0$ and $\sigma$.

\subsection*{Stochastic stimulation}

In all cases, the external current is given by Equation~\ref{e1}.
There, $\xi(t)$ is an Ornstein-Uhlenbeck process of zero mean and
unit variance, with a correlation time $\tau$. In
Figures~\ref{f4}-\ref{f6} and \ref{f8}-\ref{f10}, the value of $I_0$
was chosen so that the average number of spikes per burst was
approximately 3.5, and the value of $\sigma$ was chosen so that the
standard deviation of the number of spikes per burst was
approximately 1.5. The value of $\tau$ was chosen smaller than all
the structure in the BTAs, but not as small as to require too
extensive simulations in order to obtain smooth BTAs.

\subsection*{Model bursting neurons}

For each type of bursting (parabolic, square-wave and elliptic),
three neuron models are employed, with varying degree of biological
detail. In all cases, with the parameters chosen for our
simulations, a positive stimulus input was required in order to
obtain bursting responses. If the stimulus was sufficiently high,
burst firing was replaced by tonic activity.

\subsection*{Minimal normal-form models}

These models were designed to represent the relevant bifurcation
in their canonical form. The equations were taken from
\citet{Izhikevich2007}, except for the dynamics of the slow
current in the elliptic case, here modified to stabilize the
subthreshold behavior, since in the original model, the slow
current diverges when the fast currents are quiescent.

The fast subsystem of parabolic and square-wave bursters is described
by a single variable $V$ evolving with the dynamics of a quadratic
integrate-and-fire model neuron. This model constitutes the
canonical form of any type I neuron \citep{Ermentrout1986}.
The equation is
\begin{eqnarray}
\frac{{\rm d}V}{{\rm d}t} = V^2 + \alpha I_{\rm burst} + I_{\rm
ext}, \label{e6}
\end{eqnarray}
where $I_{\rm burst}$ is the current produced by the slow
subsystem as detailed below, and $I_{\rm ext}$ is the external
current. The coupling constant $\alpha$ is varied in
Figure~\ref{f7}, and set equal to unity in all other plots. There
is a resetting rule supplementing Eq.~\ref{e6}, stating that
whenever $V$ increases above $V_{\rm TH}$, it is discontinuously
changed to a reset value $V_{R}$, and a spike is fired.

The fast subsystem of elliptic bursters involves two variables. The
equations are usually given in radial coordinates $(r, \vartheta)$,
such that the variable representing the voltage is $r \cos
\vartheta$. The dynamics is governed by
\begin{eqnarray}
\frac{{\rm d}r}{{\rm d}t} &=& I_{\rm burst} \ r + c r^3 + d r^5 \nonumber \\
\frac{{\rm d}\vartheta}{{\rm d}t} &=& 1 \label{e7}
\end{eqnarray}
The system of Eqs.~\ref{e7} is the canonical form of a type II
neuron with a subcritical bifurcation at firing onset \citep{Brown2004}. The external current $I_{\rm ext}$ is incorporated as an
additive term in the equation for the voltage, obtained by
transforming Eq.~\ref{e7} to cartesian coordinates.

We now describe the dynamics of the slow subsystems. For parabolic
bursters, $I_{\rm burst} = u_1 - u_2$, where
\begin{eqnarray}
\frac{{\rm d}u_1}{{\rm d}t} &=& -\mu_1 u_1 + d_1 \delta(t - t_{\rm
spike}) , \nonumber \\
\frac{{\rm d}u_2}{{\rm d}t} &=& -\mu_2 u_2 + d_2 \delta(t - t_{\rm
spike}),
\end{eqnarray}
where $t_{\rm spike}$ represents the time when the $V$ crosses the
threshold voltage $V_{\rm TH}$ and generates a spike. At those
times, the slow variables $u_1$ and $u_2$ are discontinuously reset
to $u_1 + d_1$ and $u_2 + d_2$, respectively. For square-wave bursters,
$I_{\rm burst} = - u_1$, where
\begin{eqnarray}
\frac{{\rm d}u_1}{{\rm d}t} = -\mu_1 u_1 + d_1 \delta(t - t_{\rm
spike})
\end{eqnarray}
When $V$ crosses the threshold $V_{\rm TH}$, the slow variable
$u_1$ is discontinuously changed to $u_1 +d_1$. For elliptic bursters,
\begin{equation}
\frac{{\rm d}I_{\rm burst}}{{\rm d}t} = - \mu_1 (I_{\rm burst} + \lambda
r^2)
\end{equation}
In the case of elliptic bursting, no discontinuous resetting rule is
used. Instead, a spike is fired whenever $r \cos \vartheta$ crosses
the threshold 0.75. Table \ref{t0} specifies the parameters used in
the simulations.

\begin{table}[h]
\begin{center}
\scalebox{0.9}{
\begin{tabular}{|l|c|c|c|} \hline
Parameter & Parabolic & Square-wave & Elliptic \\
& bursting & bursting & bursting \\ \hline $I_0$ & -0.1 & -0.1 & 0
\\ \hline $\sigma$ & 0.25 ${\rm ms}^{1/2}$ & 1.4 ${\rm ms}^{1/2}$ & 0.75 ${\rm ms}^{1/2}$ \\ \hline
$\tau$ & 1 ms & 0.5 ms & 0.2 ms \\ \hline $V_{TH}$ & 20 & 10 & -
\\ \hline $V_{\rm R}$ & -1 & 1 & - \\ \hline $\mu_1$ & 0.1 ${\rm ms}^{-1}$ & 0.015 ${\rm ms}^{-1}$ &
0.0025 ${\rm ms}^{-1}$ \\ \hline $d_1$ & 1.1 & 0.22 & - \\ \hline $\mu_2$ & 0.02 ${\rm ms}^{-1}$ & - & - \\
\hline $d_2$ & 0.55 & - & - \\ \hline $\lambda$ & - & - & 1.25 \\
\hline $c$ & - & - & 0.4 \\ \hline $d$ & - & - & -0.2 \\ \hline {\rm
d}t & 0.1 ${\rm ms}$ & 0.05 ${\rm ms}$ & 0.01 ${\rm ms}$ \\ \hline
$\alpha$ & 1 & 1 & 1 \\ \hline
\end{tabular}}
\end{center}
\caption{\label{t0} Parameters of the minimal normal-form models
used in Figs.~\ref{f4}-\ref{f6} and \ref{f8}-\ref{f10}.
Figure~\ref{f7} was built with the same parameters except for the
coupling strength $\alpha$.}
\end{table}

\subsection*{Minimal conductance-based models}

These models are taken from \citet{Izhikevich2007}. Spiking is
induced by the fast subsystem, containing a fast persistent sodium
current and a potassium current. Bursting is modulated by the slow
subsystem, involving one or two additional slow variables. For all
bursting types, the equations of the fast subsystem read
\begin{eqnarray}
C \frac{{\rm d}V}{{\rm d}t} &=& - g_{\rm Na} \ m_\infty(V) \ (V -
V_{\rm Na}) - g_{\rm K} \ n \ (V - V_{\rm K}) \nonumber \\ & & -
g_{\rm L} \ (V - V_{\rm L}) + \alpha I_{\rm burst} + I_{\rm ext}
\nonumber \\
\frac{{\rm d}n}{{\rm d}t} &=& \left[n_\infty(V) - n\right] /
\tau_{\rm n}.
\end{eqnarray}
In Figure~\ref{f1}, $\alpha$
takes the values 0.5, 1, and 1.1. Throughout the rest of the paper,
$\alpha$ is maintained fixed at 1. The activation curves $x_\infty$
are
\begin{equation}
x_\infty (V) = \frac{1}{1 + \exp\left[(V_{1/2}^x - V)/k^x
\right]}, \label{e2}
\end{equation}
and the parameters depend on the type of bursting, as specified in
Table \ref{t1}. In the parabolic burster, the slow subsystem is
described by two variables, associated with sodium and potassium
currents, respectively. That is, the bursting current reads
\begin{equation}
I_{\rm burst} = -g_{{\rm ms}} m_{{\rm s}} (V - V_{{\rm Na}}) -
g_{{\rm ns}} n_{{\rm s}} (V - V_{{\rm K}}),
\end{equation}
and
\begin{eqnarray}
\frac{{\rm d}m_{\rm s}}{{\rm d}t} &=& \left[m_{{\rm s} \
\infty}(V) -
m_{\rm s}\right] / \tau_{\rm ms}, \nonumber \\
\frac{{\rm d}n_{\rm s}}{{\rm d}t} &=& \left[n_{{\rm s} \ \infty}(V)
- n_{\rm s}\right] / \tau_{\rm ns}. \label{e4}
\end{eqnarray}
In the square-wave and the elliptic bursters, the slow subsystem is
described by a single variable, associated with a potassium current,
\begin{equation}
I_{\rm burst} = - g_{{\rm ns}} n_{{\rm s}} (V - V_{{\rm K}}), \ \
\end{equation}
and $n_{\rm s}$ is governed by Eq.~\ref{e4}. Parameters are given in
Table \ref{t1}.

\begin{table}[h]
\begin{center}
\scalebox{0.9}{
\begin{tabular}{|l|c|c|c|} \hline
Parameter & Parabolic & Square-wave & Elliptic \\
& bursting & bursting & bursting \\ \hline $I_0$ & 1.2 $\mu$A/cm$^2$ & 4 $\mu$A/cm$^2$ & 48 $\mu$A/cm$^2$ \\
\hline $\sigma$ & 2 $\mu$A/cm$^2$ & 2.75 $\mu$A/cm$^2$ & 3 $\mu$A/cm$^2$ \\ \hline $\tau$ & 1 ms & 0.5 ms & 1 ms \\
\hline
$C$ & 1 $\mu$F/cm$^2$ & 1 $\mu$F/cm$^2$ & 1 $\mu$F/cm$^2$ \\
\hline $V_{\rm Na}$ & 60 mV & 60 mV & 60 mV \\
\hline $V_{\rm K}$ & -90 mV & -90 mV & -90 mV \\
\hline $V_{\rm L}$ & -80 mV & -80 mV & -80 mV \\
\hline $g_{\rm Na}$ & 20 mS/cm$^2$ & 20 mS/cm$^2$ & 4 mS/cm$^2$ \\
\hline $g_{\rm K}$ & 10 mS/cm$^2$ & 9 mS/cm$^2$ & 4mS/cm$^2$ \\
\hline $g_{\rm L}$ & 8 mS/cm$^2$ & 8 mS/cm$^2$ & 1 mS/cm$^2$ \\
\hline $V_{1/2}^{\rm m}$ & -20 mV & -20 mV & -30 mV\\
\hline $k^{\rm m}$ & 15 mV & 15 mV & 7 mV\\ \hline $V_{1/2}^{\rm
n}$ & -25 mV & -25 mV & -45 mV\\ \hline $k^{\rm n}$ & 5 mV & 5 mV
& 5 mV
\\ \hline $\tau_{\rm n}$ & 1 ms & 0.152 ms & 1 ms \\ \hline$g_{\rm ms}$ & 3
mS/cm$^2$ & - & - \\ \hline $V_{1/2}^{\rm ms}$ & -40 mV & - & - \\
\hline
$k^{\rm ms}$ & 5 mV & - & - \\ \hline $\tau_{\rm ms}$ & 20 ms & - & - \\
\hline $g_{\rm ns}$ & 20 mS/cm$^2$ & 5 mS/cm$^2$ & 1.5 mS/cm$^2$
\\ \hline $V_{1/2}^{\rm ns}$ & -20 mV & -20 mV & -20 mV \\ \hline
$k^{\rm ns}$ & 5 mV & 5 mV & 5 mV \\ \hline $\tau_{\rm ns}$ & 50
ms & 20 ms & 60 ms \\ \hline
\end{tabular}}
\end{center}
\caption{\label{t1} Parameters of the minimal conductance-based
models used in Figures~\ref{f4}-\ref{f6} and \ref{f8}-\ref{f10}.
Figures \ref{f1}-\ref{f3} were constructed with the same parameters,
except for $I_0$ and $\sigma$.}
\end{table}

\subsection*{Detailed conductance-based models}

The Chay-Cook model is a detailed conductance-based model of a
pancreatic $\beta$-cell that can produce parabolic, square-wave or
elliptic bursting, depending on the parameters \citep{Chay1988}. The
model includes two inward Ca$^{2+}$ currents ($I_{{\rm I}}$ and
$I_{{\rm S}}$), an outward delayed-rectifier Potassium current
($I_{{\rm K}}$), and a leakage current ($I_{{\rm L}}$).
\begin{eqnarray}
C \frac{{\rm d} V} {{\rm d} t} &=& - I_{{\rm Ca}}(V, s) - I_{\rm
K} (V, n) - I_{\rm L}(V) \nonumber \\
\frac{{\rm d}n} {{\rm d}t} &=& \lambda \frac{n_{\infty}(V) - n}
{\tau_n(V)} \nonumber \\
\frac{{\rm d}s} {{\rm d}t} &=& \frac{s_{\infty}(V, c) - s}
{\tau_s(V, c)} \nonumber \\
\frac{{\rm d}c} {{\rm d}t} &=& f\left[-\alpha I_{{\rm Ca}}(V, s) -
k_{\rm c} c\right]
\end{eqnarray}
where $V$ is the membrane potential, $c$ is the intracellular free
Ca$^{2+}$ concentration, $n$ and $s$ are the activation variables.
The ionic currents are given by
\begin{eqnarray}
I_{{\rm Ca}}(V, s) &=& I_{{\rm I}}(V) + I_{{\rm S}}(V, s) \nonumber \\
I_{{\rm I}}(V) &=& g_{{\rm I}} \ m_{\infty}(V) \ \left(V - V_{{\rm
Ca}}\right) \nonumber \\
I_{{\rm S}}(V, s) &=& g_{{\rm S}} \ s \ \left(V - V_{{\rm
Ca}}\right) \nonumber \\
I_{{\rm K}}(V, n) &=& g_{{\rm K}} \ n \ \left(V - V_{{\rm
K}}\right) \nonumber \\
I_{{\rm L}}(V) &=& g_{{\rm L}} \ \left(V - V_{{\rm L}}\right)
\end{eqnarray}
The steady-state functions and activation times are
\begin{eqnarray}
x_{\infty}(V) &=& \frac{1}{1+\exp \left[\left (
V_{x}-V\right)/S_{x}\right]}, \ \ \ {\rm for} \ x = m, n \nonumber \\
s_{\infty}(V, c) &=& \frac{1}{1 + \exp\left[2A(V, c)\right]}, \nonumber \\
\tau_{\rm n}(V) &=& \frac{\tau_{\rm n0}} {1 + \exp\left[\left(V -
V_{\rm n}\right) / S_{\rm n}\right]}, \nonumber \\
\tau_{\rm s}(V, c) &=& \frac{\tau_{\rm s0}}{2 \cosh(A(V, c))}, \nonumber \\
A(V, c) &=& \frac{V_{{\rm s}} + S_{{\rm s}} \log(\tilde{c}) -
V}{2S_{{\rm s}}},
\end{eqnarray}
and $\tilde{c} = c / (1 \mu {\rm M})$.

\begin{table*}
\begin{center}
\begin{tabular}{|l|c|c|c|} \hline
Parameter & Parabolic & Square-wave & Elliptic \\
& bursting & bursting & bursting \\
\hline $I_0$ & -50 $\mu$A/cm$^2$ & -100 $\mu$A/cm$^2$ & -50 $\mu$A/cm$^2$ \\
\hline $\sigma$ & 50 $\mu$A/cm$^2$ & 200 $\mu$A/cm$^2$ & 60 $\mu$A/cm$^2$ \\
\hline $\tau$ & 50 ms & 50 ms & 50 ms\\
\hline $\tau_n^0$ & 9.09 ms & 9.09 ms & 9.09 ms \\
\hline $\tau_s^0$ & 100 ms & 0.1 ms & 0.1 ms \\
\hline $C$ & 4524 f{}F & 4524 f{}F & 1 4524 f{}F \\
\hline $V_{\rm Ca}$ & 100 mV & 100 mV & 100 mV \\
\hline $V_{\rm K}$& -80 mV & -80 mV & -80 mV \\
\hline $V_{\rm L}$ & -60 mV & -60 mV & -60 mV \\
\hline $g_{\rm I}$ & 250 pS & 250 pS & 250 pS \\
\hline $g_{\rm K}$ & 1300 pS & 1300 pS & 1300 pS \\
\hline $g_{\rm L}$ & 50 pS & 50 pS & 50 pS \\
\hline $g_{\rm S}$ & 10 pS & 10 pS & 10 pS \\
\hline $V_{\rm m}$ & -22 mV & -22 mV & -22 mV\\
\hline $V_{\rm n}$ & -9 mV & -9 mV & -9 mV\\
\hline $V_{\rm s}$ & -22 mV & -22 mV & -22 mV\\
\hline $S_{\rm m}$ & 7.5 mV & 7.5 mV & 7.5 mV\\
\hline $S_{\rm n}$ & 10 mV & 10 mV & 10 mV\\
\hline $S_{\rm s}$ & 10 mV & 10 mV & 10 mV\\
\hline $\alpha_{\rm n}$ & 5.727 $10^{-6}$ fA$^{-1}\mu$M ms$^{-1}$ & 5.727 $10^{-6}$ fA$^{-1}\mu$M ms$^{-1}$ & 5.727 $10^{-6}$ fA$^{-1}\mu$M ms$^{-1}$\\
\hline $k_{\rm c}$ & 0.03 ms$^{-1}$ & 0.027 ms$^{-1}$ & 0.022 ms$^{-1}$\\
\hline $\lambda$ & 0.6 & 0.95 & 0.1\\
\hline $f$ & 0.0015 & 0.002 & 0.002\\
 \hline
\end{tabular}
\end{center}
\caption{\label{t3} Parameters of the Chay-Cook model models used in
Figures.~\ref{f4}-\ref{f6} and \ref{f8}-\ref{f10}.}
\end{table*}

\subsection*{Covariance analysis}

In order to disclose the stimulus features responsible for burst
generation, we use spike-triggered covariance techniques
\citep{Paninski2003, samengo2013}. We define $P_n[t_0|s(t)]$ as
the probability of generating a burst of $n$ spikes at time $t_0$
conditional to a time-dependent stimulus $s(t)$. We assume that
$P_n$ only depends on the stimulus $s(t)$ through a few relevant
features $f^1(t - t_0), f^2(t - t_0), \dots f^k (t - t_0)$. The
stimulus and the relevant features are continuous functions of
time. For computational purposes, we represent them as vectors
${\bf s}$ and ${\bf f}_i$ of $N$ components, where each component
$s_j = s(j \ \delta t)$ and $f^i_j = f^i(j \ \delta t)$ is the
value of the stimulus evaluated at discrete intervals $\delta t$.
If $\delta t$ is small compared with the relevant timescales of
the models, the discretized signal is still a good approximation
of the original continuous function. The relevant features ${\bf
f}^1, \dots , {\bf f}^k$ lie in the space spanned by those
eigenvectors of the matrix
\begin{equation}
{\rm M} = {\rm C_p}^{-1} \ {\rm C}, \label{e8}
\end{equation}
whose eigenvalues are significantly different from unity
\citep{samengo2013}. Here, ${\rm C}$ is the $N �\times N$
$n$-burst-triggered covariance matrix,
\[
({\rm C})_{i j} = \frac{1}{N_{\rm n} - 1} \sum_{t_0} s(t_i + t_0)
\ s(t_j + t_0) - s_0(t_i) \ s_0(t_j),
\]
where the sum is taken over all the times $t_0$ where bursts of
$n$ spikes are initiated, $N_{\rm n}$ is the total number of
bursts witn $n$ spikes, and $s_0(t)$ is the n-burst-triggered
average (BTA):
\[
s_0(t) = \frac{1}{N_{\rm n}} \sum_{t_0} s(t + t_0).
\]
In Equation~\ref{e8}$, {\rm C_p}$ is the $N \times N$ prior
covariance matrix,
\[({\rm C_p})_{i j} = \overline{s(t_i + t) \ s(t_j + t)} -
\left[\overline{s(t)}\right]^2,
\]
where the horizontal line represents a temporal average on the
variable $t$.

The eigenvalues of $M$ that are larger than 1 correspond to
directions in stimulus space where the stimulus segments
associated with burst generation have an increased variance, as
compared to the raw collection of stimulus segments. Such
directions are often found in cells excited by stimuli that are
large in modulus, and whose sign is irrelevant. Eigenvalues that
lie significantly below unity are associated with stimulus
directions of decreased variance. Such stimuli are often found in
cells that respond to precisely tuned values of the stimulus. In
general terms, the obtained eigenvalue is a measure of the ratio
of variances of the two ensembles. That is, an eigenvalue that is
noticeably smaller than unity indicates a certain feature for
which the ratio of the variance of the burst-triggering stimuli
and the variance of the raw stimuli is significantly small.


\bibliographystyle{apalike}


\end{document}